\documentclass[doublecol]{epl2} 

\usepackage{latexsym,amssymb,amsmath}
\usepackage{color}
\usepackage{cite}

\graphicspath{{./}{../submit_EPL/}}

\title{Dynamics and Rheology of Vesicle Suspensions in Wall-Bounded
Shear Flow}

\author{Antonio Lamura\inst{1} \thanks{E-mail: \email{a.lamura@ba.iac.cnr.it}}
\and Gerhard Gompper\inst{2} \thanks{E-mail: \email{g.gompper@fz-juelich.de}}}
\shortauthor{A. Lamura \etal}

\institute{                    
  \inst{1} Istituto Applicazioni Calcolo, CNR,
Via Amendola 122/D, 70126 Bari, Italy\\
  \inst{2} Theoretical Soft Matter and Biophysics, 
Institute of Complex Systems, Forschungszentrum 
J\"{u}lich, 52428 J\"{u}lich, Germany
}

\abstract{
The dynamics and rheology of suspensions of fluid vesicles or red blood
cells is investigated by a combination of molecular dynamics and
mesoscale hydrodynamics simulations in two dimensions. The
vesicle suspension is confined between two no-slip walls, which are
driven externally to generate a shear flow with shear rate $\dot\gamma$.
The flow behavior is studied as a function of $\dot\gamma$, the volume fraction
of vesicles, and the viscosity contrast between inside and outside fluids.
Results are obtained for the encounter and interactions of two vesicles,
the intrinsic viscosity of the suspension, and the cell-free layer near
the walls.
}

\begin{document}

\maketitle

\section{Introduction}

Suspensions of mesoscale particles in viscous liquids are ubiquitous,
with examples in biological systems (blood flow), home products (paints), 
food products (emulsions), and industrial processing (pastes).
The suspended particles can be spheres, rods, fibers, flexible and semiflexible 
macromolecules, droplets, capsules, vesicles and cells.
While the dynamics of rigid particles in suspension and their rheological
behavior have been investigated in considerable detail and are 
by now reasonably well understood \cite{mewi12}, 
much less is known about the dynamics and rheology of {\em deformable} 
particles, 
in particular in the semi-dilute regime, where hydrodynamic and steric 
interactions 
between the particles become important. 

The dynamics of soft objects, in particular under flow, depends on the physical
origin of their deformability, like the surface tension at constant volume for 
droplets, the membrane bending rigidity at fixed volume and surface area for 
vesicles, and in addition the membrane shear elasticity for capsules and cells.
Therefore, these systems have to be investigated independently to understand the
relation between the elasticity of the particles and the rheological behavior of
their suspensions.

In the dilute regime, the vesicle dynamics shows tank-treading 
(TT), tumbling (TU) and vacillating-breathing dynamics, depending on 
shear rate $\dot\gamma$ and viscosity contrast $\lambda$ 
\cite{kell82,kant05,kant06,misb06,gg:gomp07c,lebe07,vlah07}.
For TT quasi-spherical 
vesicles in three dimensions (3D), the viscosity of a dilute
suspension has been predicted to be  
\cite{dank07a,dank07b}
\begin{equation}
\label{eq:eta_danker}
\eta/\eta_{out} = 1 + \frac{5}{2} \, \phi 
                          \left[1 - \frac{\Delta}{40\pi}(23\lambda + 32) 
\right] 
\end{equation}
as a function of excess area $\Delta = 4 \pi [\frac{A}{4 \pi} 
(\frac{4 \pi}{3 V})^{2/3}-1]$ and viscosity contrast 
$\lambda=\eta_{in}/\eta_{out}$, where  
$A$ and $V$ are the surface and volume of the vesicle, 
$\eta_{in}$ and $\eta_{out}$ are the fluid
viscosities of the inner and outer fluids, respectively,  
and $\phi$ is the vesicle volume fraction.
Thus, the intrinsic viscosity $\eta_I = (\eta-\eta_{out})/(\eta_{out}\phi)$
is predicted to be a decreasing function of $\Delta$ and $\lambda$. 
Furthermore, $\eta_I$ is foreseen to have a cusp-like minimum  
at the tank-treading to tumbling (or tank-treading to vacillating-breathing) 
transition, and then to increase again with increasing $\lambda$ 
\cite{dank07a,dank07b}.
This latter behavior has been also found in the numerical calculations of 
a two-dimensional vesicle by the boundary-integral approach \cite{ghig10}. 

These theoretical predictions have been tested experimentally 
\cite{vitk08,kant08}.
While a decrease of $\eta_I$ with increasing $\lambda$ was found in 
ref.~\cite{vitk08},
in good agreement with the theoretical prediction (\ref{eq:eta_danker}), 
in contrast
an increase of $\eta_I$ was found in ref.~\cite{kant08}. 
However, the available experimental results
are not conclusive for several reasons. First,
vesicle sizes in suspensions are typically polydisperse. Second, viscosity
measurements require a minimum volume fraction $\phi$ of vesicles, 
typically 5\% to 10\%, 
and are therefore difficult to extrapolate to the dilute limit \cite{kant08}.  
Indeed, experiments have been performed recently 
\cite{leva12} which demonstrate that 
vesicle interactions become relevant for the viscosity for $\phi$ around
10\%. 

Therefore, we study here the rheology of vesicle suspensions in 
the ``semi-dilute" 
regime, where particle interactions are important, but particles are not yet 
densely packed into a glassy state. Our results are obtained from mesoscale
hydrodynamics simulations of two-dimensional (2D) model systems,  
which allow the study of larger system sizes and longer time scales.
Compared to the theory of ref.~\cite{dank07a,dank07b} and the simulations of 
ref.~\cite{ghig10}, our model includes
thermal fluctuations and has the capability of studying systems over a  
wide range of vesicle concentrations.
Our main results concern the dependence of the intrinsic viscosity on 
viscosity contrast, shear-thinning behavior, displacements
and angular oscillations in two-vesicle collisions, and the dependence of 
cell-free-layer thickness on shear rate $\dot\gamma$.

\begin{figure}
\onefigure[scale=0.42]{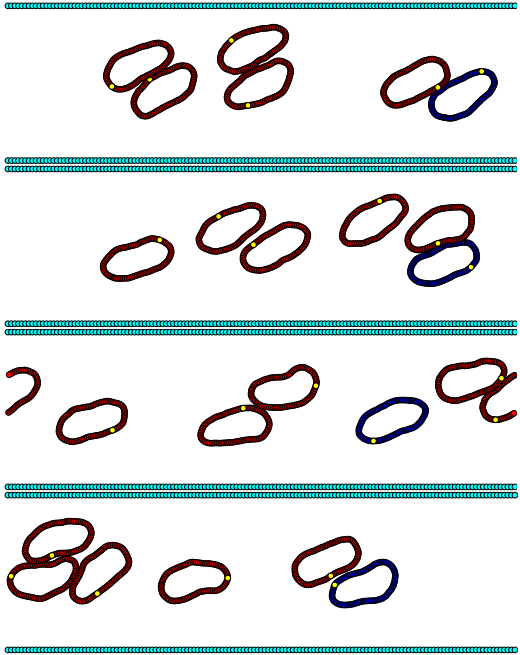}
\caption{
Configurations at consecutive times 
$\dot\gamma t = 424$, $440$, $456$, $472$ 
(from top to bottom)
of vesicles with viscosity contrast $\lambda=1.0$,
reduced area $A^{*}=0.8$, reduced shear rate $\gamma^{*}=2.0$,
and concentration $\phi= 0.14$. One vesicle is colored blue for better
visualisation of its evolution during tank-treading. 
See also movie S1 for $\lambda=2.0$, $A^{*}=0.8$, $\gamma^{*}=2.0$,
and $\phi= 0.28$.
}
\label{fig:configs}
\end{figure}

\section{Method and Model}

Each vesicle in two dimensions 
is modeled as a chain of $N_p$ beads of mass $m_p$, connected successively
in a closed ring \cite{gg:gomp08d}, see fig.~\ref{fig:configs}. 
Neighboring beads are connected to each other
by an harmonic potential with spring constant $k_h$ and
average bond length $r_0$; this keeps the perimeter length of the membrane
constant, both locally and globally.
Shapes and fluctuations are then controlled by a bending potential
$V_b=(\kappa/r_0) \sum_i (1-\cos \beta_i)$, where $\beta_i$ is the angle between
the two bond vectors at bead $i$, and $\kappa$ is the bending rigidity. 
Finally, in order to
keep the vesicle area $A$ close to the target value $A_0$, a potential
$V_A = k_A (A-A_0)^2 / (2r_0^4)$ is employed, where $k_A$ is the   
compression modulus.
Different vesicles repel each other at short distances via a shifted 
Lennard-Jones potential, which is truncated at its minimum $r_{cut}$.
Newton's equations of motions for the beads are integrated by using
the velocity-Verlet algorithm with time step $\Delta t_p$ \cite{alle92}. 

The fluid is described by multi-particle collision (MPC) dynamics, a 
particle-based mesoscale simulation technique \cite{male99,kapr08,gg:gomp09a}.
The two-dimensional fluid consists of 
$N_s$ point particles of mass $m$, whose positions ${\bf r}_i(t)$ and
velocities ${\bf v}_i(t)$, $i=1,2,...,N_s$, are continuous variables.
The evolution occurs in discrete time intervals $\Delta t_s$, and
proceeds in two consecutive steps: streaming and collision. In the 
streaming step, particles move ballistically.
In the collision step, the particles are sorted into the cells of a 
regular square lattice of mesh size $a$; all particles within each cell 
collide and exchange momentum. We employ here a variant of MPC, 
denoted as MPC-AT+a,
which conserves both linear and angular momentum locally 
\cite{gg:gomp07b,gg:gomp07h} and keeps the temperature constant 
\cite{gg:gomp07b}.
The viscosity of the MPC-AT+a fluid in two dimensions
is given by
\begin{equation}
\eta=\frac{m}{\Delta t_s} 
 \Big [  \Big ( \frac{l}{a}  \Big )^2  \Big ( \frac{n^2}{n-1}-\frac{n}{2}  
\Big ) 
+ \frac{1}{24}  \Big (n - \frac{7}{5}  \Big )  \Big ]
\end{equation}
with $l=\Delta t_s \sqrt{k_B T /m}$ the mean-free path, $k_B T$ the 
thermal energy, and $n$ the average number of particles per cell 
\cite{gg:gomp08f}.
The system of size $L_x \times L_y$ is placed between two
horizontal walls which slide along the $x$ direction
with velocities $v_{wall}$ and $-v_{wall}$, respectively.
Periodic boundary conditions are used along the $x$ direction. A 
bounce-back rule with virtual particles ensures no-slip boundary conditions 
at the walls \cite{gg:gomp01g,gg:gomp07h}. 
This generates a linear flow profile $v_x=\dot\gamma y$ with shear
rate $\dot\gamma= 2 v_{wall} / L_y$. 

To describe the fluid-membrane interaction, membrane beads are modeled 
as hard disks, see fig.~\ref{fig:configs}.  
The radius $r_p$ of the disks is chosen large enough to ensure mutual overlap 
and a complete coverage of the membrane to prevent fluid 
particles from crossing the membrane.
Since it is very important to conserve linear and angular 
momentum for vesicles with viscosity contrast 
\cite{gg:gomp07h,gg:gomp09g}, we employ the following  
scattering rule between fluid particles and membrane disks. Scattering
occurs only when a fluid particle $j$ and a membrane disk $i$
overlap and move towards each other, so that 
the conditions $|{\bf r}_i-{\bf r}_j| < r_p$ and 
$ ({\bf r}_i-{\bf r}_j) \cdot ({\bf v}_i-{\bf v}_j) < 0$ are satisfied. 
A second disk $k=i\pm 1$ in the same membrane, with
$\displaystyle \min_{k=i\pm1} |{\bf r}_k-{\bf r}_j|$, is selected to perform
a three-body collision which conserves linear and angular momenta 
\cite{gg:gomp09g}.
The MPC collision step is then performed only for those fluid 
particles which did not participate in the membrane scattering, in order 
to avoid multiple collisions with the same disk in subsequent time steps.
The fluids in the interior and exterior of the vesicle
may differ in their particle mass to control viscosity.
Membrane disks interact with walls via bounce-back.

In experiments with vesicles in shear flow, inertial effects are negligible
since the Reynolds number $Re=\dot\gamma \rho R_0^2 / \eta_{out}$, where 
$\rho=n m/a^2$ is the fluid mass density, is typically very small.
We express the results in dimensionless quantities, such as
the reduced area $A^*=A_0/\pi R_0^2$ (where
$R_0 = L_0/(2 \pi)$ is the mean vesicle radius with membrane 
length $L_0$) and the reduced
shear rate $\gamma^*=\dot\gamma \eta_{out} R_0^3/\kappa$. 
We set $n=10$, $l_{out}=0.0064 a$ with 
$l_{in}=l_{out} \sqrt{m_{out}/m_{in}}$ (in the following the subscripts
$out/in$ will refer to quantities outside/inside of the vesicle).
This implies that the viscosity contrast is
$\lambda=\eta_{in}/\eta_{out} \simeq m_{in}/m_{out}$.
We use the system size $L_x= 18.95 R_0$, $L_y=5.79 R_0$, 
mean radius $R_0=7.6 a$, and $v_{wall}$ such that
$Re < 0.2$ for all the cases we considered with 
$0.4 \le \gamma^* \leq 10.0$. Finally, we set
$m_{in}$ such that $0.1 \leq \lambda \leq 13.0$, $m_p=3 m_{out}$, 
$N_p=480$, $\Delta t_p = \Delta t_s / 64$, $r_p=r_0=a/10$, $r_{cut}=a$,
$\kappa=6.58 k_B T R_0$, 
$k_A= 4 \times 10^{-4} k_B T$, $k_h=3 \times 10^2 k_B T$, 
and $A_0$ such that $0.8 \leq A^* \leq 0.95$.
This value of $\kappa$ gives rise to a similar amplitude
of undulation modes as for lipid bilayer membranes in 3D (where 
$\kappa_{3D} \simeq 10 k_BT$).
With these choices for $k_A$ and $k_h$, the area and the length of the
vesicle are kept constant with a deviation less than $1\%$ of the target 
values for all simulated systems.

\begin{figure}
\onefigure[scale=0.41]{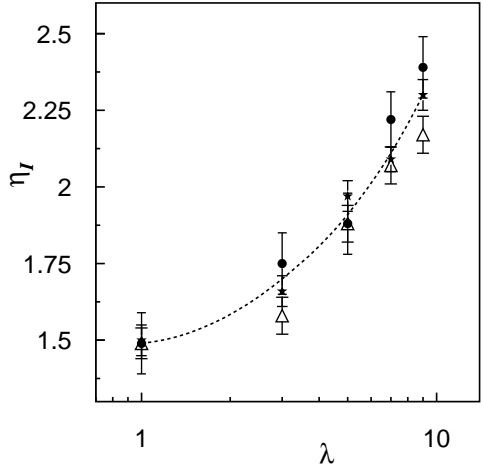}
\caption{The intrinsic viscosity 
$\eta_I=(\eta-\eta_{out})/(\eta_{out} \phi$) 
as a function of the viscosity contrast $\lambda$
for reduced shear rate $\gamma^{*}=2.0$,
reduced area $A^{*}=0.8$, and concentrations
$\phi= 0.05$ ($\bullet$), $0.09$ ($\triangle$), and $0.14$ ($\star$). 
The dashed line is the interpolation to the
data ($\star$).  The tank-treading-to-tumbling 
transition occurs at $\lambda_c\simeq 3.7$ for $A^{*}=0.8$ 
in the KS theory \cite{kell82}.
}
\label{fig.visc}
\end{figure}

\section{Results}
\subsection{Suspension Viscosity}
We first consider dilute and semi-dilute
monodisperse suspensions of vesicles with fixed reduced shear rate 
$\gamma^{*}=2.0$.
Systems with $N_V=2$, $4$, or $6$ vesicles are studied, 
corresponding to concentrations $\phi= 0.05$, $0.09$, and $0.14$, 
for different viscosity contrasts $\lambda$.
A few typical vesicle configurations for $\phi= 0.14$ and $\lambda=1.0$ 
are displayed in fig.~\ref{fig:configs}.
The suspension viscosity $\eta$ is calculated numerically 
\cite{gg:gomp08c} from the component $\sigma_{xy}$ of the stress tensor, 
so that $\eta=\sigma_{xy}/\dot\gamma$ \cite{mewi12}.

\begin{figure}
\onefigure[scale=0.41]{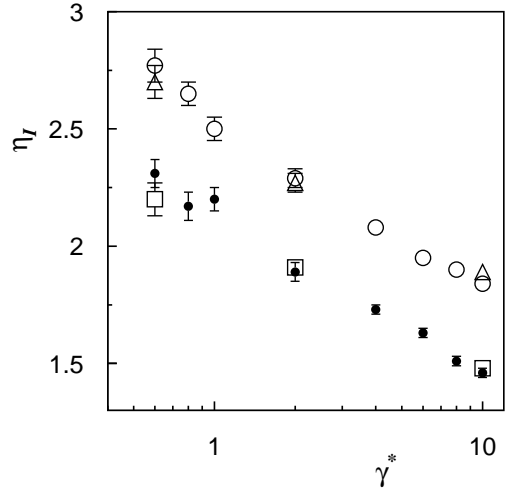}
\caption{
The intrinsic viscosity 
$\eta_I=(\eta-\eta_{out})/(\eta_{out} \phi)$ 
as a function of the reduced shear rate $\gamma^{*}$
for reduced area $A^{*}=0.8$, concentration 
$\phi= 0.28$, and viscosity contrasts 
$\lambda= 2.0$ $(\bullet)$ and $5.0$ $(\circ)$
with $L_x \times L_y = (18.95\times 5.79)R_0$, and 
$\lambda= 2.0$ $(\Box)$ and $5.0$ $(\triangle)$
with $L_x \times L_y = ( 15.79 \times 6.84)R_0$. 
When not visible, error bars are comparable
with symbols size.
}
\label{fig.visc12}
\end{figure}

The relative viscosity 
$\displaystyle (\eta-\eta_{out})/\eta_{out}$ 
is a linear function of $\phi$ for $A^{*}=0.8$ and various values of $\lambda$
as predicted by the Einstein relation
\cite{belz81}.
In fig.~\ref{fig.visc}, the intrinsic viscosity $\eta_I$ 
is shown as a function of $\lambda$ for various concentrations with 
$A^{*}=0.8$. 
An increase of $\eta_I$ with the viscosity contrast is observed. 
We do not find an indication of a non-monotonic behavior ---
as predicted theoretically by eq.~(\ref{eq:eta_danker}) 
in refs.~\cite{dank07a,dank07b} for quasi-spherical vesicles in 
3D and obtained numerically in 2D in ref.~\cite{ghig10}  ---
in the explored range $1.0 \leq \lambda \leq 9.0$ of viscosity contrasts,
although the dynamic behavior changes from TT to 
TU at intermediate values of $\lambda$.   
In two dimensions, the TT-to-TU transition is predicted to occur 
at $\lambda_c \simeq 3.7$ for $A^*=0.8$ in the Keller-Skalak (KS)
theory \cite{kell82}. However, 
thermal vesicle undulations, which are neglected in KS theory, 
produce a continuous crossover from TT to TU for bending rigidities around
$\kappa = 6.4 k_BT R_0$ \cite{gg:gomp09g}, with  
$A^*=0.7$ and $\gamma^* \lesssim 6$.
Thus, our simulation results of increasing $\eta_I(\lambda)$ are in 
qualitative agreement with the 
experimental results of ref.~\cite{kant08} for semi-dilute systems.

We consider next the behavior of monodisperse
concentrated suspensions with $\phi=0.28$ (12 vesicles)
with reduced area $A^*=0.8$ for viscosity contrasts $\lambda=2.0$, $5.0$
as a function of the reduced shear rate $\gamma^*$.
Two systems of size $L_x\times L_y = (18.95 \times 5.79) R_0$ and 
$(15.79 \times 6.84)R_0$ are investigated, which
have the same area but the latter being a vesicle radius $R_0$ wider 
than the former.
The results of the intrinsic viscosity $\eta_I$ are displayed in 
fig.~\ref{fig.visc12}. 
The values of $\eta_I$ are not affected by the system size.
For both values of $\lambda$, a significant 
shear-thinning behavior is found  over more than one decade in the reduced 
shear rate. The data also show that it is difficult to reach the low shear-rate
plateau in simulations.  
This is due to the importance of thermal motion at low shear rates,
but may also be related to the broad TT-to-TU transition in 2D 
\cite{gg:gomp09g} where some tumbling events already appear in the TT regime.
This shear-thinning is mainly due to the formation of cell-free layers near 
the walls, as expected from the F{\aa}hraeus-Lindqvist effect \cite{fahr31}. 
The formation of cell-free layers will be discussed in detail below. 
An analysis of the effective viscosity in the central part of the channel,
as derived from the local shear rate, shows that shear-thinning of the
core region, as observed in bulk red blood cell suspensions in 3D, both 
experimentally \cite{skal81} and in simulations \cite{gg:gomp11k}, is
not significant in 2D in the considered concentration range.

\begin{figure}
\onefigure[scale=0.41]{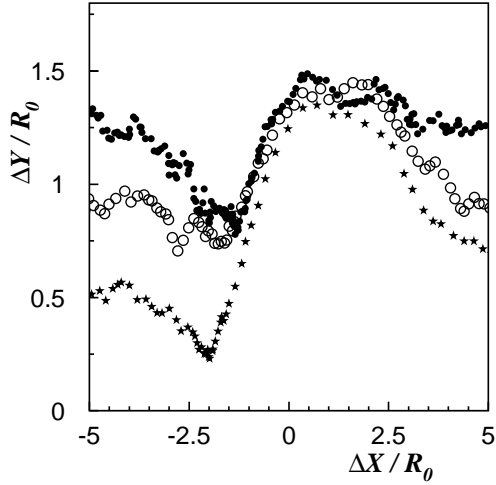}
\caption{
Relative vertical displacement of the centers of mass $\Delta Y/R_0$
of two scattering vesicles 
with respect to the relative horizontal distance $\Delta X/R_0$
for $\lambda=1.0$, $A^{*}= 0.8$,
and shear rates 
$\gamma^{*}= 2.0$ ($\bullet$), $5.0$ ($\circ$), and $10.0$ ($\star$).
}
\label{fig.cm}
\end{figure}

\begin{figure}
\onefigure[scale=0.41]{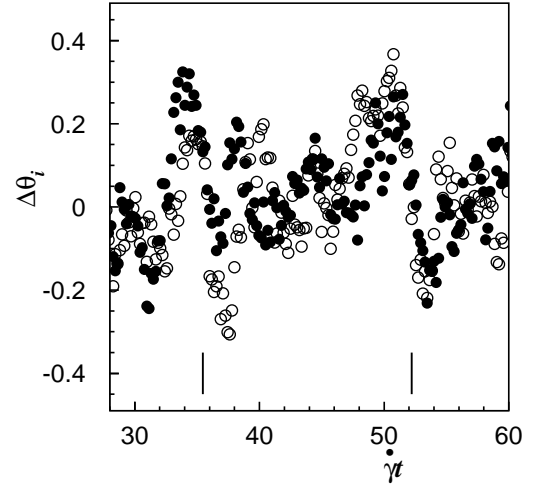}
\caption{
Dynamics of two vesicles over two subsequent interaction events.
Deviations $\Delta \theta_i$
of the inclination angles of both the two vesicles ($i=1,2$, 
indicated by the two types of symbols)
from the average stationary value are shown as a function of time
for the run in fig.~\ref{fig.cm}  with $\gamma^*=2.0$.
The vertical lines denote the times of the closest relative distance
between the two vesicles. See also movie S2, which displays vesicle
interactions in the time range $30 < \dot\gamma t < 60$. 
}
\label{fig.thetabis}
\end{figure}

\begin{figure}
\onefigure[scale=0.41]{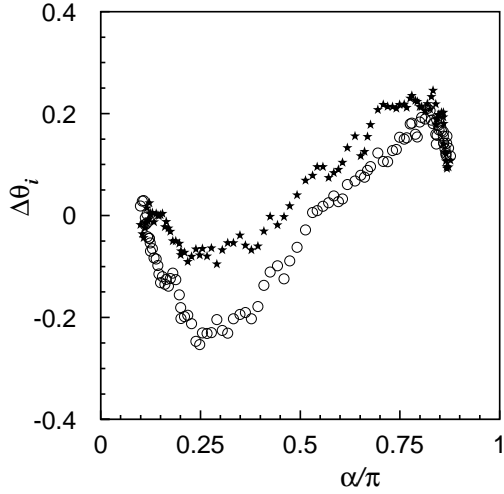}
\caption{
Deviations $\Delta \theta_i$ of the inclination angles of two interacting
vesicles ($i=1,2$, indicated by the two types of symbols) from the 
average stationary value as a function of $\alpha$, the angle between 
the direction connecting the vesicles centers of mass and the
flow direction, for the run in fig.~\ref{fig.cm} with $\gamma^*=2.0$. 
Data are averaged over four subsequent interaction events.
}
\label{fig.thetater}
\end{figure}

\subsection{Vesicle Interactions}
Following the experimental work in refs.~\cite{kant08,leva12}, we study 
the interaction
between two vesicles with $A^{*}=0.8$ in the TT regime ($\lambda=1.0$). 
Figure \ref{fig.cm} displays the relative vertical displacement 
of the centers of mass $\Delta Y=y_{cm1}-y_{cm2}$ during scattering with 
respect to the horizontal displacement $\Delta X=x_{cm1}-x_{cm2}$,
where $(x_{cm1},y_{cm1})$ and $(x_{cm2},y_{cm2})$ are the positions of 
the centers of mass of the two vesicles,
for three different shear rates $\gamma^{*}=2.0$, $5.0$, and $10.0$. 
Movie S2 illustrates the vesicle interaction for $\gamma^{*}=2.0$.
Figure \ref{fig.cm} shows four important effects: (i) When the vesicles are 
released
from their initial positions in the upper and lower halves of the channel, they
migrate towards the center due to the wall-induced lift force 
($\Delta X/R_0 \lesssim -2$);
(ii) upon collision, the vesicles are displaced and reach a maximum 
in their vertical separation $\Delta Y$ corresponding to
the small vesicle diameter ($\Delta X/R_0 \simeq 0$); (iii) immediately after 
the collision,
the vertical displacement is larger than before the collision 
($\Delta X/R_0 \simeq 2.5$);
(iv) the vesicle continue to migrate towards the center line 
($\Delta X/R_0 \gtrsim 2.5$).
Different shear rates mainly determine the migration rate, but seem to have 
little
effect on the collision process itself.  
Good agreement for the collision process is found with experimental results,
which are obtained for much wider channels (see fig.~2 of ref.~\cite{kant08}). 
The increased vertical separation after scattering is in qualitative 
agreement with recent theoretical predictions for quasi-spherical vesicles 
and large inter-vesicle distances \cite{gires2012}. 

The interaction process can also monitored in time by considering the behavior
of the distance $d$ between the centers of mass of the vesicles and the
deviations $\Delta \theta_i = \theta_i-\theta_0$ of the inclination angle
$\theta_i$ $(i=1,2)$ 
of the two vesicles from its average stationary value $\theta_0$. 
This latter is shown in 
fig.~\ref{fig.thetabis} for two consecutive scattering events for the same
run of fig.~\ref{fig.cm} with reduced shear rate $\gamma^*=2.0$. We observed
a correlation of the tilt angles of the vesicles 
when the relative distance is at minimum, in agreement with the experimental
results (compare with fig.~8 of ref.~\cite{leva12}).
In fig.~\ref{fig.thetater}, the
deviations  $\Delta \theta_i$ are shown as a function of the 
relative displacement angle $\alpha$, 
defined as the angle between the direction along the 
vesicles centers of mass and the
flow direction, for the run in fig.~\ref{fig.cm} with $\gamma^*=2.0$. 
Data are averaged over four subsequent scattering processes. 
The maximum and
minimum of $\Delta \theta_i$ occur approximately at $\alpha \simeq 3\pi/4$ and 
$\alpha \simeq \pi/4$, corresponding to the compression and stretching 
directions of the shear flow field, respectively. 
This is again in agreement
with the recent experimental results (see fig.~6 of ref.~\cite{leva12}),
except for a small displacement of the minimum position. 

\subsection{Cell-Free Boundary Layer}

\begin{figure}
\onefigure[scale=0.41]{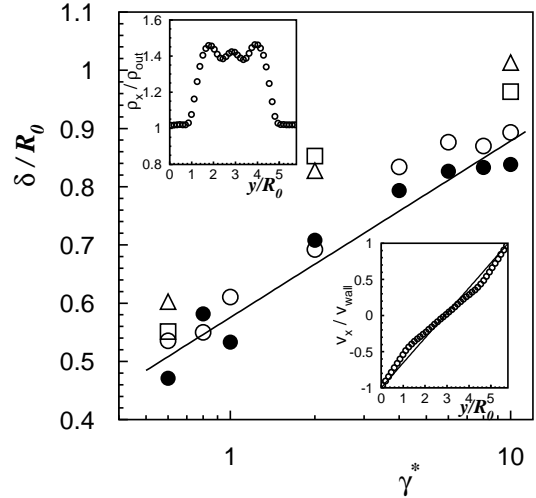}
\caption{
The ratio of the 
average thickness $\delta$ of the vesicle-free boundary layers
to the vesicle radius $R_0$
as a function of the reduced shear rate $\gamma^{*}$
for reduced area $A^{*}=0.8$, concentration 
$\phi= 0.28$, and viscosity contrasts 
$\lambda= 2.0$ $(\bullet)$ and $5.0$ $(\circ)$
with $L_x \times L_y = (18.95 \times 5.79)R_0$, and 
$\lambda= 2.0$ $(\Box)$ and $5.0$ $(\triangle)$
with $L_x \times L_y = (15.79\times 6.84)R_0$. 
Error bars are comparable with symbols size. 
The full line is a logarithmic fit to data points ($\bullet$).
Insets: (left) Fluid mass density and 
(right) fluid velocity component $v_x$, averaged along the flow ($x$) 
direction,  
both across the channel with $\lambda=2.0$ and $\gamma^*=10.0$.
(Right) The full line corresponds to the imposed shear flow profile.
}
\label{fig.delta}
\end{figure}

For red blood cells (RBCs) in capillary flow, it was first found by 
F{\aa}hraeus and Lindqvist \cite{fahr31} that cells are depleted from a 
layer near the vessel wall. It is now well understood that this is a 
consequence of the wall-induced hydrodynamic lift force on the cells
\cite{gg:gomp09g}. 
For concentrated systems with $\phi=0.28$,  we have measured the 
average thickness $\delta$
of the cell-free boundary layers near the walls. $\delta$ is 
defined as the time average of $(d_1(t)+d_2(t))/2$
with $d_{1,2}(t)=\min_{i=1,...,N_V} l_i(1,2)(t)$, where
$N_V$ is the number of vesicles and $l_i(1,2)$ the closest
distance of $i$-th vesicle membrane from either of the two walls.
The results are reported in fig.~\ref{fig.delta} 
as a function of the reduced shear rate.
The ratio $\delta/R_0$ increases with $\gamma^*$ for both the values
of the considered viscosity contrasts. The data are consistent  
with a logarithmic growth of $\delta$ with increasing $\gamma^*$.
With increasing system width, the values of $\delta$ grow, as also observed
in two-dimensional simulations of RBC-like vesicles in capillary flow 
\cite{bagc07}.

The existence of boundary layers is also supported by considering
the fluid mass density 
profiles in the steady state averaged along
the flow direction $x$.  
The mass density is lower in the cell-free boundary layers at the walls  
due to the absence of vesicles with a heavier fluid inside
(see the inset of fig.~\ref{fig.delta}).
This effect of the boundary layer is also evident in 
steady-state velocity profiles, which display a smaller effective shear 
rate in the center and a higher shear rate near the wall, as compared to the
imposed shear rate $\dot\gamma$ (see the inset of 
fig.~\ref{fig.delta}).

\begin{figure}
\onefigure[scale=0.41]{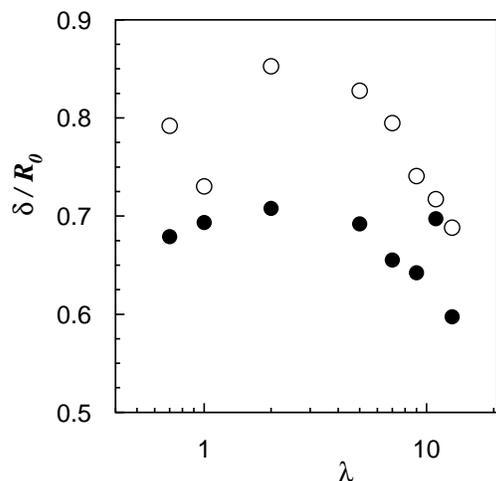}
\caption{
The ratio of the 
average thickness $\delta$ of the vesicle-free boundary layers
to the vesicle radius $R_0$
as a function of the viscosity contrast $\lambda$
for reduced area $A^{*}=0.8$, reduced shear rate $\gamma^{*}=2.0$, 
concentration $\phi= 0.28$, and system sizes  
$L_x \times L_y = (18.95 \times 5.79)R_0$ ($\bullet$) and 
$L_x \times L_y =  (15.79\times 6.84)R_0$ ($\circ$).  
Error bars are comparable with symbols size.
}
\label{fig.deltalambda}
\end{figure}

Finally, the behavior of $\delta$ as a function of
 $\lambda$ for $\gamma^{*}=2.0$ is
shown in fig.~\ref{fig.deltalambda} for the two system sizes. 
It is evident that there is a pronounced non-linear 
dependence of $\delta$ on $\lambda$, with the boundary layer decreasing
at high values of the viscosity contrast, attaining a maximum near the
TT-to-TU transition. This behavior can be related to the dependence of the
lift force on the viscosity contrast, which has been shown \cite{gg:gomp09g}
to be a decreasing function of $\lambda$. 
In the TT phase, the increase of $\delta$ might be due to the reduction of the 
tilt angle with increasing $\lambda$, which facilitates the sliding of vesicles 
past each other and allows them to be squeezed more easily into the
center of the channel. 
In the TU phase, tumbling of vesicles is suppressed near the wall, which 
further reduces the lift force \cite{gg:gomp09g}. 

The cell-free layer thickness $\delta$ is found to grow with $L_y$, in 
agreement with
simulations of RBCs in cylindrical microchannels in 3D (compare 
fig.~10 of ref.~\cite{fedo10b}).  $\delta$ is found to increase with 
$\dot\gamma$,
in agreement with the results of RBC simulations in 3D presented in 
fig.~8 of ref.~\cite{freu11}, but at odds with the simulation results 
in fig.~11 of ref.~\cite{fedo10b} and experimental results of 
ref.~\cite{kim07}. 
This apparent discrepancy can be partially resolved by taking a closer
look at the investigated range of shear rates.
In our 2D case, the increase of $\delta$ occurs for $\gamma^* \lesssim 7$; 
in ref.~\cite{freu11}, it is seen for $\gamma^* \lesssim 20$ in 3D; and
in ref.~\cite{fedo10b}, $\delta$ is found to slightly decrease for 
$\gamma^* \gtrsim 4$ 
(in refs.~\cite{freu11,fedo10b} the average shear rate is used,
computed from the average velocity in a Poiseuille flow;
for RBCs in three dimensions, we use $\tau=\eta_0 R_0^3/\kappa$
with mean radius $R_0=3.4 \mu$m, bending rigidity $\kappa=50 k_BT$, and the
plasma viscosity $\eta_0=0.0012$ Pa s --- 
which implies $\tau=0.22$ s --- to determine the dimensionless shear rate).
Although it is of course difficult to compare results of 2D and 3D systems
quantitatively, we can conclude that there is a critical reduced shear rate 
$\gamma^*_c \simeq 10$ 
below which $\delta$ is increasing with $\gamma^*$, and above which
$\delta$ is constant or slowly decreasing.
The value of $\gamma^*_c$ depends, of course, on the channel width
and the vesicle volume fraction \cite{fedo10b}; the value above should 
be valid for volume fractions around 0.3 and channel widths of about 
three vesicle diameters.

\section{Summary and Conclusions}

We have investigated the dynamical behavior of semi-dilute suspensions of 
vesicles or red blood 
cells under shear flow in a narrow gap between two walls. The advantage of the
Couette geometry compared to Poiseuille flow is that wall effects and effects
of a non-linear flow profile do not interfere. 
We find the intrinsic viscosity to increase monotonically with increasing 
viscosity contrast, a pronounced shear-thinning behavior with increasing 
shear rate
due to the F{\aa}hraeus-Lindqvist effect, displacements
and angular oscillations in two-vesicle collisions in good agreement 
with experiments, 
and an increase of the cell-free-layer thickness with shear rate $\dot\gamma$
below a critical reduced shear rate $\gamma^*_c \simeq 10$.

\acknowledgments
Fruitful discussions with D. Fedosov, I. O. G\"{o}tze, S. Messlinger, 
H. Noguchi, and M. Peltomaeki are gratefully acknowledged.

\bibliographystyle{eplbib}
\bibliography{amphiphile,gompper}

\end{document}